\documentstyle[sprocl]{article}

\def\Journal#1#2#3#4{{#1} {\bf #2}, #3 (#4)}

\def\PLB{{\em Phys. Lett.}  B}

\def\PRC{{\em Phys. Rev.} C}

\def\ra{\rightarrow}

\def\be{\begin{equation}}
\def\ee{\end{equation}}
\def\bea{\begin{eqnarray}}
\def\eea{\end{eqnarray}}

\begin{document}
\begin{flushright}
JLAB--96--17\\ July 1996
\end{flushright}
\bigskip

\title{FORM FACTORS OF KAON SEMILEPTONIC DECAYS \footnote{Talk 
presented by A.A. at
PANIC'96, Willimsburg, VA, May 22--28, 1996; to be published in the 
Proceedings}}
\author{Andrei Afanasev$^{a,b}$ and W.W. Buck$^b$}
\address{$^a$ Thomas Jefferson National Accelerator Facility\\
12000 Jefferson Ave., Newport News, VA 23606, USA}
\address{$^b$ The NuHEP Research Center, Department of Physics\\
Hampton University, Hampton, VA 23668, USA}

\maketitle
\abstracts{A calculation of the semi--leptonic decays of the kaon ($K_{l3}$) 
is presented. The
results are direct predictions of a covariant model of the pion and kaon
introduced earlier by Ito, Buck, Gross. The weak form factors  
for $K_{l3}$ are predicted with absolutely no parameter adjustments
of the model. We obtained for the form factor parameters: 
$f_-(q^2=m_l^2)/f_+(q^2=m_l^2)=-0.28$  and $\lambda_+$= 0.028,
both within experimental error bars. Connections of this approach to heavy
quark symmetry will also be discussed.}

A covariant model based on coupled Bethe--Salpeter and 
Schwinger--Dyson equations with an NJL--like interaction   
was successfully used earlier to describe
electromagnetic properties of pions and kaons.\cite{ibg,bwi}

Using  this model,
we analysed form factors of $K_{l3}$ decays (describing the 
processes $K\ra\pi\l\nu$). Assuming that only the vector weak current
contributes to the decays, the matrix element may be written as
$$ J_\mu={G_F \sin \theta_C\over\sqrt{2}}[f_+(t)(P_K+P_\pi)_\mu+
f_-(t)(P_K-P_\pi)_\mu],$$
where $P_K$ and $P_\pi$ are the four--momenta of the $K$ and $\pi$
mesons, $G_F$ is a Fermi coupling constant, $\theta_C$ is a Cabibbo angle,
and $f_+$ and $f_-$ are dimensionless form factors which can
depend only on $t=(P_K-P_\pi)^2$.  For the decay channel, the physical
region is limited by the masses of the lepton, pion and kaon:
$m_l^2\leq t\leq (m_K-m_\pi)^2.$

In our model, the pion and the kaon both share the same light quark mass 
of 250 MeV.   It is this mass,
together with the corresponding mesonic vertex functions (wave functions) 
employed in the
calculation of the elastic and transition form factors of the pion and kaon
that is now employed in the $K_{l3}$  decays addressed here.   The pion 
calculation provides a  non-perturbative
solution to the axial anomaly for space--like $Q^2$ and the charged pion 
form factor \cite{ibg}
while the kaon wave function provides a good description of
the charged and neutral kaon form factors.\cite{bwi}   

The main results of the predictive calculation 
of the $K_{l3}$ 
decays are illustrated in the Table 1 and compared to selected theoretical 
models and experiment (see also Ref.\cite{ab}). 
Results are in quite good agreement 
with available empirical data for all
observables studied and, when one quark is infinitely heavy, we reproduce the 
heavy quark symmetry limit for the ratio $f_-/f_+$.\cite{hqs}

\medskip
{\small Table 1: Model predictions for the parameters of $K_{l3}$ 
decay form factors. 

$^*$From the corresponding values of $\lambda_+$ and
$\lambda_0$; $^{**}$From the corresponding values of 
$\lambda_+$ and $\xi_A(0)$.}
\smallskip

\centerline{\begin{tabular}{|l|c|c|c|c|c|} \hline
&CPT & VMD \cite{review} & ISGW2 \cite{QM}& This work&  Experiment
\cite{pdg} \\ \hline
$\lambda_+$&0.031 \cite{cpt} & 0.0245 &
0.019 & 0.028& 0.0286$\pm$0.0022 ($K_{e3}$)\\ 
 &0.0328 \cite{shabalin} & & & &\\ \hline
$\xi_A$& --0.164$\pm$0.047$^*$ \cite{cpt}& --0.28 
& --0.28&--0.28&--0.35$\pm$0.15 
($K_{\mu 3}$) \\ 
& --0.235 \cite{shabalin}& & & &\\ \hline
$\lambda_0$&0.017$\pm$0.004 \cite{cpt}& 0.0 &
--0.005$^{**}$ & 0.0026& 0.004$\pm$0.007 ($K^+_{\mu 3}$) \\ 
&0.0128 \cite{shabalin}& & & & 0.025$\pm$0.006 ($K^0_{\mu 3}$) \\ \hline
\end{tabular}}
\medskip

It should be noted that Jefferson Lab measurements of $\pi$ charge form 
factor (E--93--021), kaon charge form factor (E--93--018), 
a $\gamma^*p\ra\gamma p$
measurement at the $\pi^0$ pole \cite{afanas94} and
$K_{l3}$ transitions,
\cite{finn}  will further constrain 
theoretical 
models addressing all
electroweak pion--kaon reactions.  More to the point, all of these experiments 
are connected; in fact, we contend that they must be connected without new 
parameter fits or adjustments.

It is speculated that to employ these 
theoretical techniques in the study of the nucleon, 
our quark
masses must be accompanied by a rather large 
inter--quark kinetic energy in order to bind the nucleon and extract form 
factors.

\section*{Acknowledgments}
The work of A.A. was partially supported by the US Department of Energy under
contract DE--AC05--84ER40150; the work of W.W.B. was partially supported by 
the National Science Foundation Grant Number HRD-9154080.

\end{document}